\documentclass[runningheads]{svmult}

\usepackage{makeidx}   
\usepackage{graphicx}  
\usepackage{subeqnar}  
\usepackage{multicol}  
\usepackage{physprbb}  
\makeindex             

\begin{document}
\title*{General Properties of Recurrent Bursts from SGRs}
\toctitle{General Properties of Recurrent Bursts from SGRs}
%
%
\titlerunning{General Properties of Recurrent Bursts from SGRs}
%
\author{R.L.Aptekar\inst{1}
\and P.S.Butterworth\inst{2}
\and T.L.Cline\inst{2}
\and D.D.Frederiks\inst{1}
\and S.V.Golenetskii\inst{1}
\and V.N.Il'inskii\inst{1}
\and E.P.Mazets\inst{1}
\and V.D.Pal'shin\inst{1}}
\authorrunning{R.L.Aptekar et al.}
%
%
\institute{Ioffe Physico-Technical Institute, St.Petersburg, 194021, Russia
\and Goddard Space Flight Center, Greenbelt, MD 20771, USA}

\maketitle              

\vspace{0.3cm}

Recurrent short gamma-ray bursts with soft energy spectra were discovered  
with the Konus experiment onboard the Venera 11 and Venera 12 missions
in March, 1979~\cite{Mazets79a}. The famous superintense gamma-ray outburst
on 1979 March 5 was followed by a series of 16 weaker 
bursts from source FXP~0526-66, over a period of several years~\cite{Mazets81,Gol84}.
Also, in March, 1979 three short bursts were detected and localized with the Konus 
experiment from another source, B1900+14~\cite{Mazets79b}.
It was suggested that repeated soft bursts 
represent a distinct class of events different in their origin from the majority of gamma-ray 
bursts~\cite{Mazets81,Mazets82}. In 1983, the Prognoz 9, ICE, and SMM spacecraft detected
a dense series of soft recurrent burst from a third source, 
1806-20~\cite{Atteia87,Laros87,Kouv87}, confirming the existence of a new class of objects. 
These sources of recurrent soft bursts were named Soft Gamma Repeaters, SGRs.
The fourth source, SGR~1627-41 was discovered and localized only
in 1998~\cite{Hurley99a,Woods99,Smith99,Mazets99a}.
In 1997 two bursts were observed coming from a fifth SGR~1801-23~\cite{Hurley99b,Cline00}. 
The burst emission from SGRs tends to be concentrated into short periods (weeks to months) 
of intense activity separated by relatively long periods (years) of quiescence. 

A Konus catalog of SGR activity was prepared with aim of enabling 
the detailed study of repeated bursts and 
comparison of the characteristics of different sources~\cite{Aptekar01}.
This catalog contains observational data on the bursting activity of all 
five known SGRs accumulated with Konus gamma-ray burst 
experiments onboard Venera 11--14, Wind and Kosmos-2326 spacecraft in the period 
1978--2000. These data on rates, time histories and energy spectra were obtained with similar 
instruments and are collected together in a comparable form.
The catalog is available electronically at http://www.ioffe.rssi.ru/LEA/SGR/Catalog/. 

\begin{figure}[t]
\vspace*{-0.2cm}
\centering\includegraphics[width=\textwidth,bb=90 300 460 415]{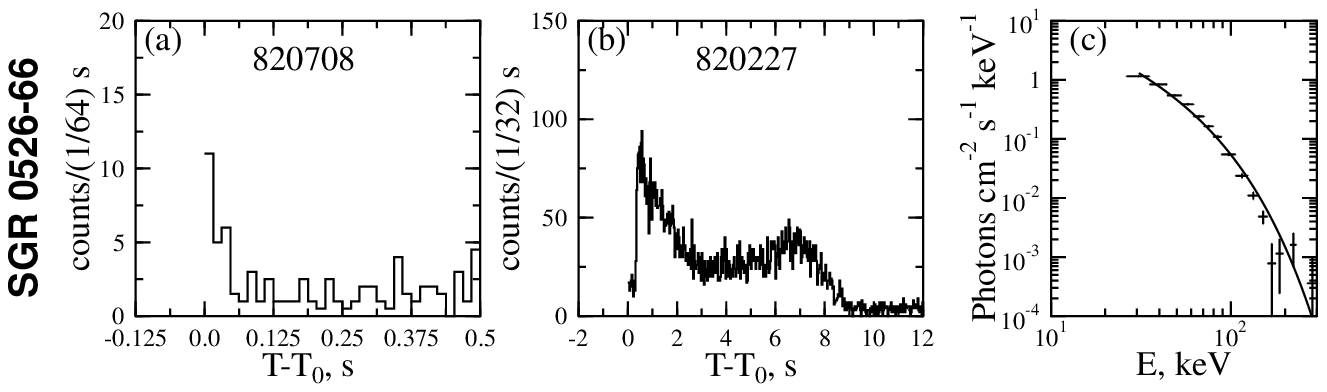}\\
\vspace*{-0.2cm}
\centering\includegraphics[width=\textwidth,bb=90 300 460 415]{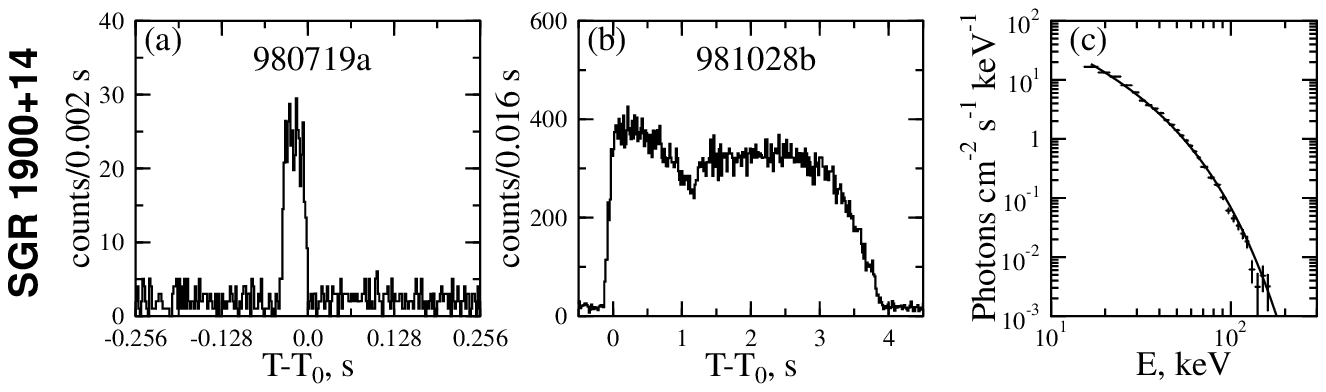}\\
\vspace*{-0.2cm}
\centering\includegraphics[width=\textwidth,bb=90 300 460 415]{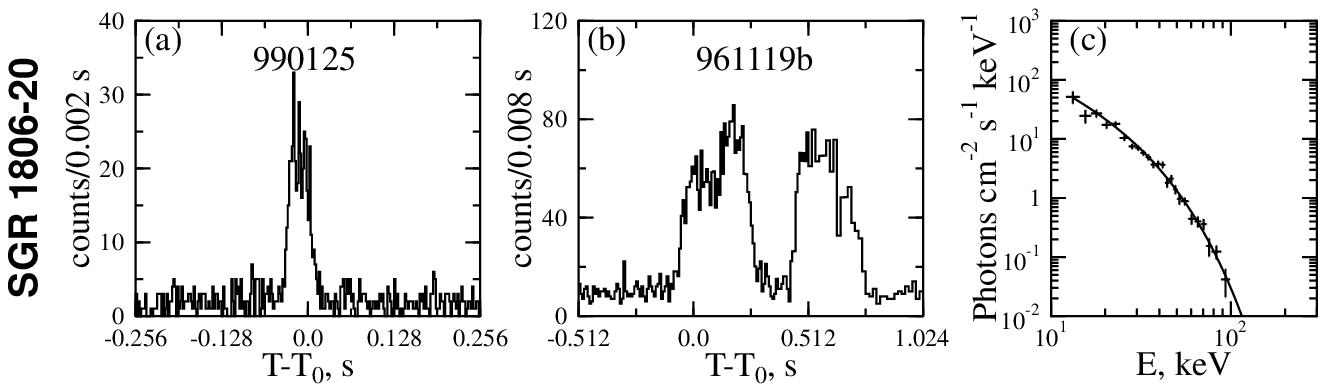}\\
\vspace*{-0.2cm}
\centering\includegraphics[width=\textwidth,bb=90 300 460 415]{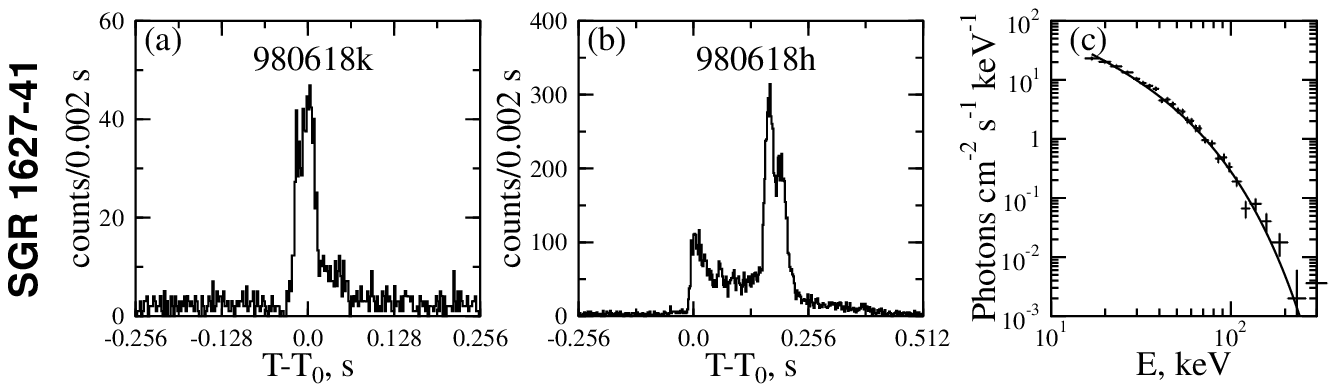}\\
\vspace{-0.4cm}
\caption{Examples of time histories and energy spectra for four SGRs 
(\textbf{a}) the shortest burst, (\textbf{b}) the longest burst, 
(\textbf{c}) the typical energy spectrum.}
\label{eps1}
\vspace*{-0.5cm}
\end{figure}

\begin{figure}[t]
\includegraphics[height=6cm,bb=80 282 240 492]{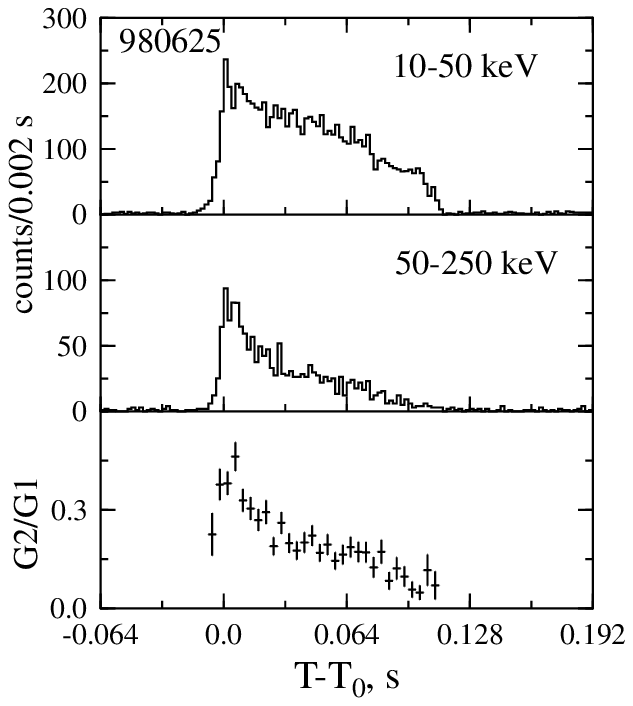}
\hfill
\includegraphics[height=6cm,bb=85 210 290 420]{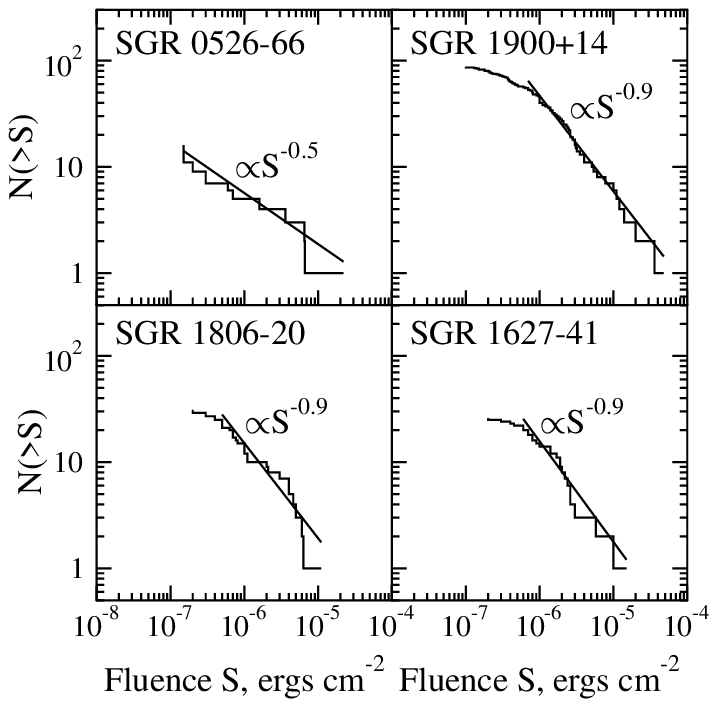}\\
\vspace*{-0.2cm}
\parbox[t]{0.55\textwidth}{\vspace{-0.5cm}\caption{An example of a burst from SGR~1627-41
with strong spectral variability.}\label{eps2}}
\hfill
\parbox[t]{0.4\textwidth}{\vspace{-0.5cm}\caption{Cumulative fluence distributions
of SGR bursts.}\label{eps3}}\\
\vspace*{-0.2cm}
\end{figure}

The catalog data illustrate all the well-known general properties of the repeated soft bursts: 
short durations, simple profile, rapid rise times ($<100$~ms) and 
soft spectra ($kT \sim 20-–30$~keV). 
As an example, the time histories and energy spectra of the SGRs are shown in 
Fig.~\ref{eps1}. Spectral variability is apparent only in the course of a number
repeated bursts of the SGR~1627-41~\cite{Mazets99a}.
Fig.~\ref{eps2} illustrates one of such bursts. 

The catalog contains various characteristics of repeated bursts such as burst durations, fluences, 
peak fluxes, hardness of energy spectra and so on. These data enable users to perform 
statistical studies of possible relations between these properties.
Some results of such investigations were 
published by Mazets et al.~\cite{Mazets99b} for SGR~1900+14, 
by Woods et al.~\cite{Woods99} for SGR~1627-41 and by G\"o\v{g}\"us et al.~\cite{Gogus99,Gogus00}
for SGR~1900+14 and SGR~1806-20. As further example of use of the catalog data,
the cumulative fluence distributions for four SGRs are shown in Fig.~\ref{eps3}.

An important feature of SGRs is the giant periodic burst.
The March 5, 1979 event remained the only one of this kind for a long time. 
Detailed study of the giant outburst on August 27, 1998
from SGR~1900+14 and a comprehensive comparison with the earlier event~\cite{Mazets99c}
showed that giant outbursts in SGRs sources are an intrinsic though rare 
stage in the evolution of these objects. Another unusual giant outburst was 
detected by the Konus-Wind experiment from SGR~1627-41 on June 18, 1998~\cite{Mazets99a}.
Its energy output and peak luminosity are close to the values of the other two giant events.
Its peculiarities include a longer rise time, no long tail and no pulsations. 

Conclusion, the catalog reveals general similarities between 
properties of the known SGRs. At the same time there are also evident distinctions. Studying 
such similarities, distinctions, and individual features should lead us to deeper understanding 
of the fleeting but extremely powerful processes operating in the SGRs.

This work was supported by Russian Aviation and Space Agency Contract, RFBR grant N 99-
02-17031 and CRDF grant RP1-2260.
\vspace{-0.3cm}

%

\end{document}